# Anatomically Guided Motion Correction for Placental IVIM Parameter Estimation with Accelerated Sampling Method

**Mbaimou Auxence NGREMMADJI[1], Freddy ODILLE[1,2], Charline BERTHOLDT[1,3], Marine BEAUMONT[1,2], Olivier MOREL[1,3] and Bailiang CHEN[1,2]**

[1]INSERM U1254, IADI, Université de Lorraine, Nancy, France
[2]CIC-IT 1433, CHRU Nancy, Vandœuvre-lès-Nancy
[3]Maternité Régionale de Nancy, France

Corresponding author: Bailiang CHEN (e-mail: bailiang.chen@inserm.fr).

**ABSTRACT** Intravoxel incoherent motion (IVIM) is a diffusion-weighted magnetic resonance imaging (MRI) method that may be applied to the placenta to help diagnose abnormal pregnancies. IVIM requires prolonged scan times, followed by a model-based estimation procedure. Maternal or fetal motion during the scan affects the accuracy of this estimation. In this work, we proposed to address this challenging motion correction and data fitting problem by using additional anatomical information that is routinely collected at the beginning of the examination. Super-resolution reconstruction (SRR) was applied to these anatomical data, to provide a patient-specific, 3D isotropic, anatomic reference. Our first contribution is a novel framework with a two-step motion correction that uses both IVIM and the SRR anatomic data, accounting for both intra- and inter-scan, non-rigid motion. Our second contribution is an automation and acceleration of the IVIM data fitting, using a state-of-the-art Bayesian-type algorithm, modified with a preconditioned Crank-Nicholson (pCN) sampling strategy. The accuracy of the IVIM parameter fitting was improved by the proposed motion correction strategy, as assessed by the mean absolute fitting error in the region of interest, which was 4.14 before and 3.02 after correction (arbitrary units of signal intensity). The novel sampling strategy accelerated parameter estimation by 39% in average, with the same accuracy as that of the conventional Bayesian approach. In conclusion, the proposed method may be applied to obtain fast and reliable IVIM parameter estimates in challenging scenarios such as prenatal MRI.

## I. INTRODUCTION

The placenta is a vital organ necessary for the exchange of nutrients and wastes between the mother and the fetus during pregnancy. Its abnormality can lead to serious pathologies such as placenta accreta spectrum disorder (PAS). It is characterized by invasion of the myometrium and leads to abnormal vascularization at the interface between the uterus and placenta. This abnormal attachment can potentially lead to fatal uncontrolled hemorrhage. The International Society of Ultrasound in Obstetrics and Gynecology (ISUOG) recommends the use of ultrasound as the first-line imaging modality for diagnosis because of its easy accessibility, safety and low cost [1]. Magnetic Resonance (MR) imaging has been promoted as an adjunct imaging tool for perinatal screening because of its larger Field of View (FoV), better imaging quality, volumetric and multidirectional nature, richer tissue contrast and higher spatial resolution. Therefore, it has great potential to aid the diagnosis of antenatal PAS. The Intravoxel Incoherent Motion (IVIM) model is a candidate MR technique for the diagnosis of PAS. Each voxel assumes two compartments that separate the intra and extravascular diffusion of the incoherent motion of water molecules within the capillaries. The IVIM parameters are as follows: (a) blood perfusion fraction, (b) blood pseudo-diffusion and (c) water molecular diffusion. One major advantage of the IVIM model in terms of placental function assessment is that it is a non-contrast MR technique for indirect perfusion measurement.

In practice, the acquisition of IVIM images is based on the conventional diffusion-weighted imaging (DWI) sequence with multiple b-values [6], [7]. This results in a long acquisition time, with the impact of motion on parameter estimation [5]. The two main challenges in in vivo placental MR are motion management and the estimation of IVIM parameters.

### A. MOTION MANAGEMENT

The major sources of motion are uterine contraction, maternal respiration, and fetal movement, which can affect IVIM parameter estimation and may potentially affect diagnosis. This motion leads to in-plane warping and through-plane deformation. Prospective motion correction methods, such as the navigator-based slice location technique [18] cannot be used because of the longer acquisition time for the placental IVIM images. We mainly focused on retrospective motion correction.

The commonly proposed IVIM motion correction methods mainly use only IVIM data. Guyader et al. [39] proposed a comprehensive pipeline based on automatic 3D non-rigid registration. They first corrected the misalignment within each single b-value 3D volume via a group-wise registration method, and then used a pairwise method to align each b-value ($b > 0$) image volume to the first b-value ($b=0$) volume because of the high signal-to-noise ratio (SNR) of the latter. Flouri et al. [19] proposed an alternating model-driven registration (MDR) technique that incorporates a motion correction technique into parameter estimation. A volume generated using the theoretical signal values was created and used as a target to guide the registration. Pairwise co-registration was performed based on free-form deformation (FFD). They demonstrated improved anatomical delineation and precision of the parameter maps. Kornaropoulos et al. [40] proposed a deformable group-wise registration framework that considers both shape and intensity changes by adding constraints from the MR diffusion behavior. They further extended this model to estimate the diffusion parameter.

In practice, the aforementioned registration methods do not consider the slice gap that is usually inserted during in vivo placental MR image acquisition. Owing to the short acquisition time and large field of view (FoV), such a gap can leverage the trade-off between acquisition time and image quality in the clinical setting. However, this causes inconsistencies in the slice direction, which is difficult to handle if only IVIM data are considered. Fortunately, in addition to functional IVIM data, the clinical placental MR protocol often starts with the acquisition of anatomical MR images in different directions, with lower through-plane spatial resolution but high in-plane resolution. These images can be used to reconstruct isotropic anatomical MR images, which provide morphological quantification and defines anatomical coordinates, that can be used as the reference space.

### B. IVIM PARAMETERS

Learning-based [42] and model-based methods have been investigated for IVIM parameter estimation. In this research, we focus on model-based methods. Many IVIM model-based parameter estimation algorithms have been proposed and compared in previous studies [9], [13]. There are two main types of model-based methods: deterministic and probabilistic methods. Typical deterministic methods include the standard full non-linear least squares (LSQ) method using, for instance, the Levenberg-Marquadt (LM) algorithm. Segmented partially linearized least-squares (SEG) and segmented doubly linearized least squares (SEGb) are examples of deterministic approaches [9]. Deterministic methods can be affected significantly by data noise. Therefore, Bayesian estimation methods have attracted considerable attention. It allows the integration of prior information to improve accuracy [10] and allows a better representation of the uncertainty in the estimates. Different information priors have been previously employed. Freiman et al. [12] used spatial homogeneity prior for the estimation of the IVIM parametric maps. Orton et al. [11] proposed a hierarchical Gaussian prior for liver IVIM data, and demonstrated its potential applicability to other organs. Bayesian estimation methods extract parameters based on the approximation of the posterior using sampling methods such as Markov chain Monte Carlo (MCMC) methods, with the random walk Metropolis-Hastings (rw) method as the most commonly used technique. However, this sampling method can be computationally expensive in a higher-dimensional space.

### C. CONTRIBUTIONS

Instead of considering parametric acquisition alone, we consider the entire MR examination as an entity, and all occur in the same patient-specific anatomical space. Therefore, we introduced an anatomically guided motion correction method. It uses the acquisition-specific anatomic coordinates through super-resolution reconstruction with motion-compensated anatomical acquisition data. Aligned IVIM data were co-registered to this anatomic reference via deformable registration. IVIM parameters were then estimated using Bayesian fitting. We propose the use of a novel MCMC method, the preconditioned Crank-Nicholson (pCN) algorithm [14] to accelerate the MCMC sampling strategy in IVIM fitting. Our innovative contributions can be summarized as follows:

- A new motion correction strategy for placental IVIM data that considers anatomical MR images has been introduced;
- A new Bayesian estimation method with improved computation time for IVIM parameters is employed.

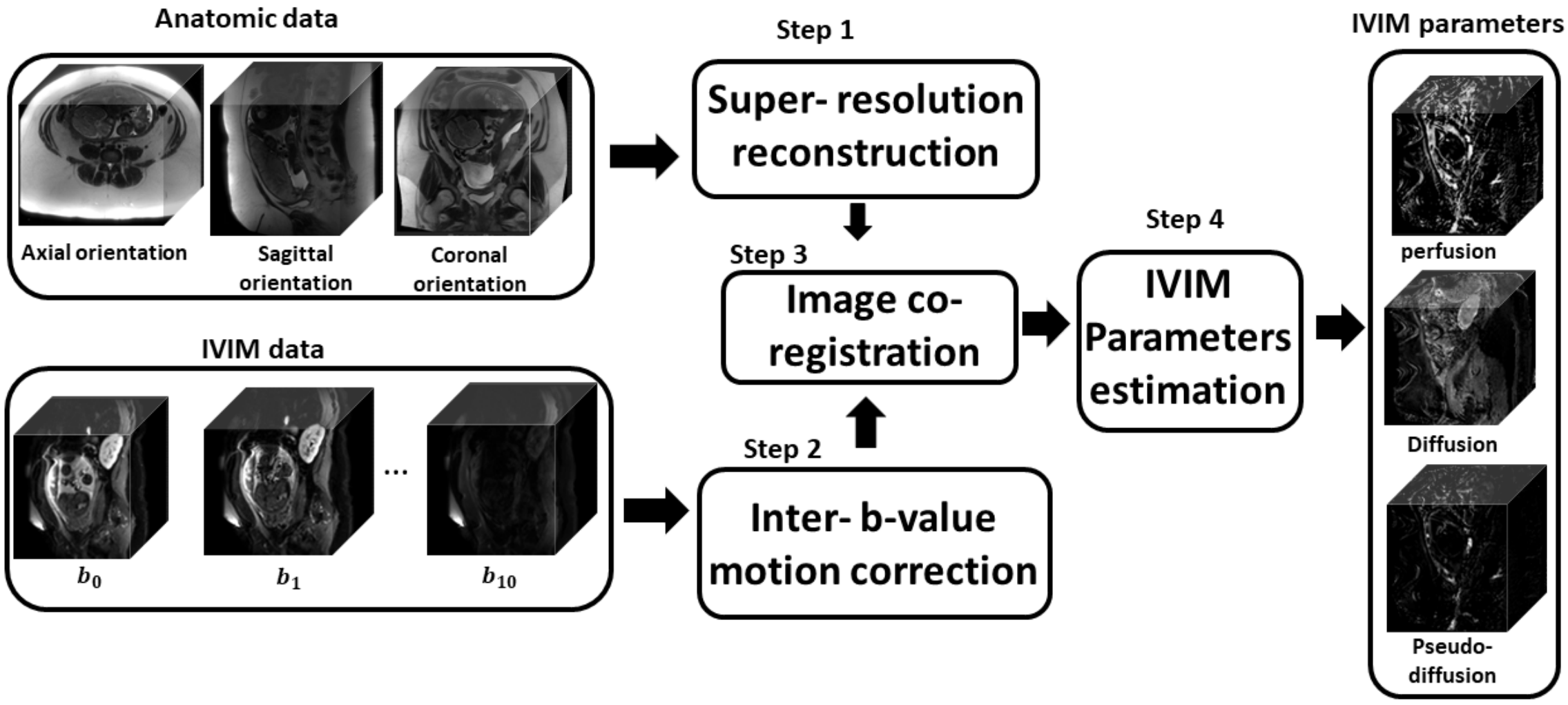

FIGURE 1. Overview of the proposed method including the new motion correction strategy and IVIM parameters estimation.

## II. PROPOSED METHOD

The main objectives of this work are: (a) the IVIM motion correction and (b) the placental IVIM parametric map estimation. In fact, in vivo MR data acquisition is impacted by motion artefacts, from the mother and the fetus. The lack of the patient-specific global reference for MR data alignment is the main constrain along with noisy parametric maps obtained when using the least-square approach. Moreover, the conventional Bayesian approach is computationally expensive.

The proposed scheme is divided in two parts: the motion correction part is performed firstly followed by the IVIM parametric map estimation. The motion correction part consists of the estimation of constructing subject-specific anatomic reference space using the motion-corrected super-resolution reconstruction method (Step 1 in Figure 1) and the IVIM data alignment using a two-step non-rigid registration (Step 2 and 3 in Figure 1). The second part consists in IVIM estimation where we have proposed a new parameter estimation method to accelerate the procedure (Step 4 in Figure 1).

### A. ESTIMATION OF THE ANATOMIC REFERENCE

Anatomic $T_2$-weighted MR images were acquired for the estimation of the anatomic reference, which was performed in two steps: (a) the alignment of $T_2$-weighted MR images and (b) super-resolution reconstruction.

#### 1) ALIGNMENT OF $T_2$-WEIGHTED MR IMAGES

Let $y_1$, $y_2$ and $y_3$ be the multi-slice anatomic $T_2$-weighted MR images acquired in the axial, coronal and sagittal planes with high in-plane spatial resolution and low through-plane spatial resolution as shown in Figure 2. Before estimating the anatomic reference, we aligned $y_1$, $y_2$ and $y_3$ using the rigid intersection-based slice-to-volume image registration method introduced in [28], and adapted in [31]. This strategy takes advantage of the intersection of acquisition planes of $y_1$, $y_2$ and $y_3$.

#### 2) SUPER-RESOLUTION RECONSTRUCTION

The motion-corrected anatomic $T_2$-weighted images denoted $\widetilde{y_1}$, $\widetilde{y_2}$ and $\widetilde{y_3}$ are then used for the estimation of the anatomic reference using the super-resolution reconstruction technique. It consists in solving the following variational problem:

$$\hat{x} \coloneqq argmin_x \sum_{i=1}^{3} || D_i B_i T_i x - \widetilde{y_i} ||^2 + \lambda_1 R_1(x) \qquad (1)$$

where $D_i$ is the downsampling operator, $B_i$ is the blurring operator and $T_i$ is the warping that takes the reconstructed image from the desired reconstructed orientation to the orientation of the acquisition of $\widetilde{y_1}$. $R_1$ is the Beltrami regularization weight, and $\lambda_1$ is its weight. $x$ is the super-resolved anatomic image and $\hat{x}$ its estimate.

To solve problem (1), we used the primal-dual algorithm [15]. Beltrami regularization was used because of its improved reconstruction quality as shown by the high sharpness index [22]. The estimated super-resolved anatomic image $\hat{x}$ is then used as the reference for IVIM registration. The latter has both high in-plane and through-plane isotropic spatial resolution, as illustrated in Figure 2.

### B. MOTION CORRECTION USING THE ANATOMIC REFERENCE

After estimating the anatomical reference, we introduce the proposed motion correction strategy for IVIM data in this section. It is a two-step method designed as follows: (a) in the first step inter-b-value motion correction is performed by aligning the data acquired with the same slice location but different b-values, and (b) in the second step, the outcome of step (a) is registered to the anatomic super-resolved reference estimated in Subsection II-A (co-registration). We used free form deformation (FFD) [23], which can correct geometric distortions in the IVIM data [21].

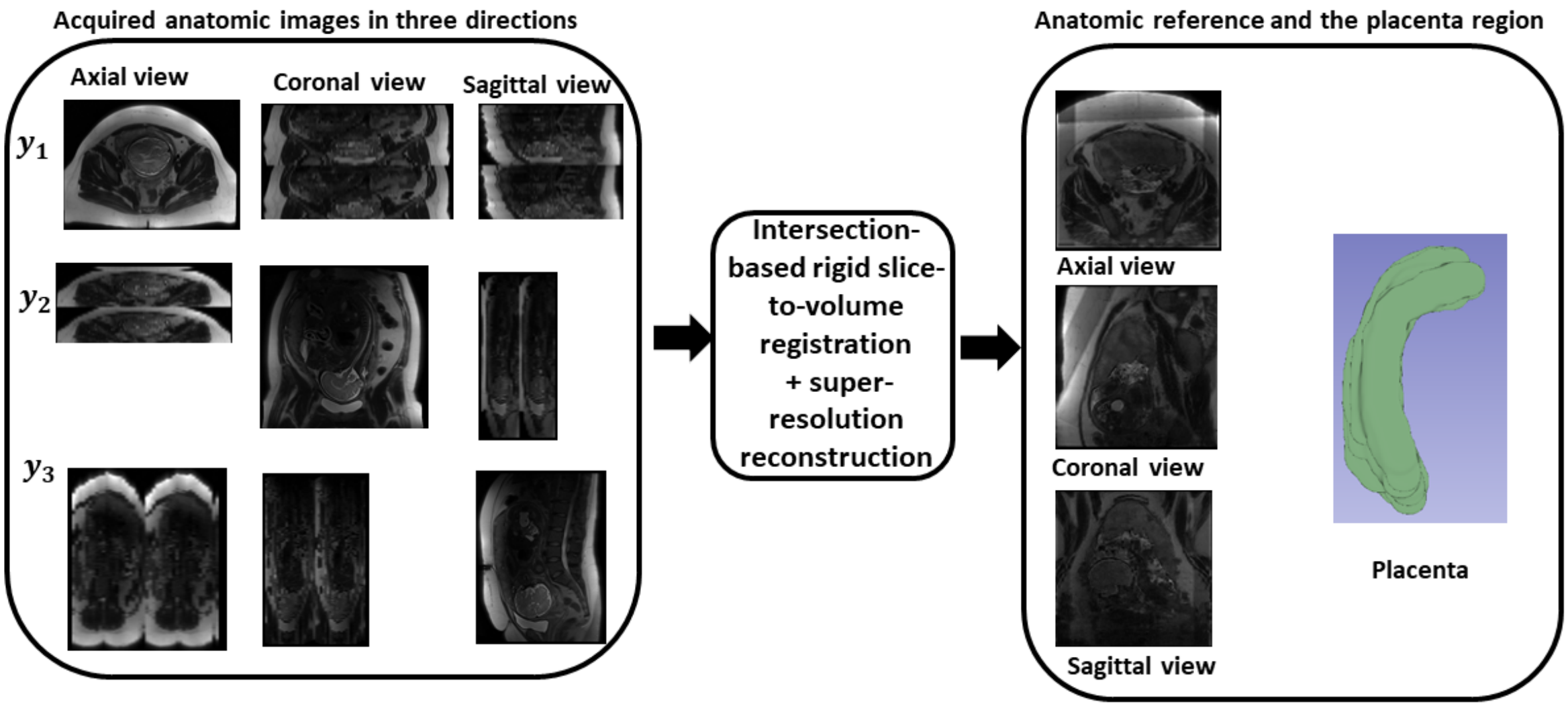

FIGURE 2. Steps for the estimation of the anatomic reference.

#### 1) INTER-B-VALUE MOTION CORRECTION

We used the deformable registration method introduced by Vishnevskiy et al. [23]. Let $\Omega \subset \mathrm{R}^2$ be the image domain, $t_i: \Omega \to \mathbb{R}^2$ the deformation of the moving IVIM slice $I_m^i: \Omega \to \mathbb{R}$ acquired using the b-value $b_i$, for $i \neq 0$, and $I_f^0: \Omega \to \mathbb{R}$ the IVIM slice obtained with the first b-value $b_0$ used as the reference. For inter-b-value motion correction, we solved the following optimization problem:

$$t_i^* \coloneqq argmin_{t_i} \mathcal{F}\left(t_i; I_m^i, I_f^0\right) + \lambda_2 R_2(t_i) \tag{2}$$

where $\mathcal{F}$ is the image dissimilarity metric, $R_2$ is the isotropic total variation (TV) [23], and $\lambda_2 > 0$ is its weight. In this study, we used the local correlation coefficient (LCC) as the image similarity. The deformation $t_i$ is parameterized by interpolation of the displacement on a regularly spaced control point grid: $t_i \coloneqq t_i(k_i)$. We used first order B-spline (linear) interpolation method. Considering this parametrization, we imposed the regularization on the control grid $k_i$ as follows:

$$k_i^* \coloneqq argmin_{k_i} \mathcal{F}\left(t_i(k_i); I_m^i, I_f^0\right) + \lambda_2 R_2(t_i(k_i)) \tag{3}$$

The limited Memory Broyden-Fletcher-Goldfarb-Shanno algorithm (LBFGS) [24] algorithm was employed to numerically solve problem (3).

#### 2) IMAGE CO-REGISTRATION USING THE ANATOMIC REFERENCE

The isotropic high-resolution anatomic reference $\hat{\mathfrak{x}}$ estimated in Subsection II-A is firstly interpolated to a grid with the same size as the IVIM data (3D) $V^0$ obtained with the first b-value $b_0$. Let $\hat{\mathfrak{x}}_{interp}^i$ be a slice of the interpolated anatomic image (reference image) and $I_{interbv}^{i,0}$ the corresponding slice in the $b_0$ image, from the inter-b-value motion-corrected IVIM data (moving image). Co-registration is performed by solving the following problem:

$$k_i^* = argmin_{k_i} \mathcal{F}\left(t_i(k_i); I_{interbv}^{i,0}, \hat{\mathfrak{x}}_{interp}^i\right) + \lambda_3 R_2(t_i(k_i)) \tag{4}$$

where $\mathcal{F}$ is the local correlation coefficient, $R_2$ is the isotropic total variation (TV) [23], and $\lambda_2$ is its weight. We also used the LBFGS [24] algorithm to solve (4).

### C. IVIM PARAMETERS ESTIMATION USING BAYESIAN METHOD

Let us remind the IVIM model:

$$y_i^{IVIM} \coloneqq G_{b_i}(f, d, ds) + \varepsilon_i \tag{5}$$

where $y_i^{IVIM}$ is the acquired signal with b-value $b_i$, for $i = 0, \ldots, N_b - 1$, $N_b$ is the number of b-values employed, and $f, d$ and $ds$ are the blood perfusion, water molecular diffusion and blood pseudo-diffusion, respectively. $\varepsilon_i$ is the noise related to the image acquisition. We assume that it is a Gaussian distribution with variance $\sigma_i^2$. The function $G_{b_i}$, a bi-exponential model, is defined as follows:

$$G_{b_i}(f, d, ds) \coloneqq y_0^{IVIM}\left(f e^{-b_i ds} + (1 - f) e^{-b_i d}\right) \tag{6}$$

where $y_0^{IVIM}$ is the signal acquired with the initial b-value $b_0$. Under physical considerations, the blood fraction perfusion $f \in [0,1]$ ($f$ is a percentage) and the blood pseudo-diffusion are faster than the water diffusion: $ds > d$. The mixing of both higher and lower b-values $b_i$ is required to consider both compartments. In the remainder of this subsection, we begin with the formulation of the Bayesian approach to explain how to integrate the preconditioned Crank-Nicholson (pCN) sampling method into IVIM parameter estimation, and demonstrate how pCN can accelerate the sampling.

### 1) LIKELIHOOD DISTRIBUTION

Assuming that noise $\varepsilon_i$ is normally distributed, the likelihood is given as follows:

$$P(y_i^{IVIM} \mid f,d,ds\, ;\, S_0,\sigma_i^2) \propto (\sigma_i^2)^{N_b/2} e^{-\frac{1}{2\sigma_i^2}\sum_{i=0}^{N_b-1}[y_i^{IVIM} - G_{b_i}(f,d,ds)]^2} \quad (7)$$

Similar to Orton et al. [11], we used the marginalization on $S_0$ and $\sigma_i^2$, leading to the following marginalized likelihood:

$$P(y \mid f,d,ds) \propto [yy^t - \frac{y^t G_b}{y^t y}]^{N_b/2} \quad (8)$$

Where $y := [y_0^{IVIM}, y_1^{IVIM}, \ldots, y_{N_b-1}^{IVIM}]^t$ and $G_b := [G_{b_0}, \ldots, G_{b_{N-1}}]^t$.

### 2) PRIOR DISTRIBUTION

Following Orton et al. [11], we use a hierarchical prior distribution over the transformed variables: $F := log(f) - log(1-f)$, $D := log(d)$ and $Ds := log(ds)$. The transformed variables are no longer constrained: $F, D, Ds \in \mathbb{R}^{NV}$, where $NV$ is the number of voxels. This facilitates the IVIM parameter estimation. We denote the variable to be estimated as $\theta := (F, D, Ds)$. The following prior distribution is considered:

$$P(\theta|\mu,\Sigma) \propto e^{-\frac{1}{2}(\theta-\mu)^t\Sigma^{-1}(\theta-\mu)} \quad (9)$$

where $\mu := (\mu_f, \mu_d, \mu_{ds})$ denotes the mean of the parameter $\theta$ and $\Sigma$ is the covariance matrix in the region of interest (ROI).

### 3) HYPERPARAMETERS DISTRIBUTION

We used a noninformative Jeffrey's prior as proposed by Orton et al. [11] to model the hyperparameters (or hyper-priors) $\mu$ and $\Sigma$ :

$$P(\mu,\Sigma) = |\Sigma|^{-1/2} \quad (10)$$

### 4) POSTERIOR DISTRIBUTION AND SAMPLING

The posterior distribution of the variables of interest $(\theta,\mu,\Sigma)$ is given by the Bayes' rule:

$$P(\theta,\mu,\Sigma \mid y) = \frac{P(\mu,\Sigma)\, P(y \mid \theta)\, P(\theta)}{P(y)} \quad (11)$$

The above posterior cannot be sampled directly because of the unknown denominator. Therefore, MCMC techniques are generally adopted for such problems which involve an exhaustive sampling method that is time consuming. In general, the proposal for the MCMC sampling technique is given by:

$$\theta^{n+1/2} := \pi(\theta^n) \quad (12)$$

Where $\theta^{n+1/2}$ is the proposed point, $\pi : \mathcal{F} \to \mathcal{E}$ is the sampler, $\mathcal{F}$ and $\mathcal{E}$ are two given spaces and $\theta^n$ is the current state of the chain. The preconditioned Crank-Nicholson (pCN) algorithm defines the following proposal:

$$\theta^{n+1/2} := \sqrt{1-\rho^2}\theta^n + \rho\delta^{n+1/2} \quad (13)$$

with $\rho \in ]0,1[$ and $\delta^{n+1/2} \sim \mathcal{N}(0, C)$ and $C$ is the covariance matrix. This proposed point is accepted with probability:

$$\alpha(\theta^{n+1/2}, \theta^n) = min\left(1, \frac{P(y|\theta^n)}{P(y|\theta^{n+1/2})}\right) \quad (14)$$

In equation (14), the hyper-priors $\mu$ and $\Sigma$ are no longer needed to be estimated because $\alpha$ depends only on $\theta$, therefore the step of ROI-based $\mu$ and $\Sigma$ estimation can be skipped. The initialization of the IVIM parameters was performed via the segmented fitting approach [30], in the same way as that performed by Orton et al [11], where an MCMC random walk was used. Algorithm 1 summarizes the proposed IVIM parameter estimation method.

**Algorithm 1: IVIM-pCN**

1. Inputs: $\rho_1, \rho_2, \rho_3 \in ]0,1[$, the covariances $C_1$, $C_2$, $C_3$ and the initial guest $\theta^1 := (F^1, D^1, Ds^1)$
2. Output: IVIM parameters estimate $\hat{\theta}$
3. For iteration n= 2, …, maxIter
   a) Sample the proposal of $F$

      $$F^{n+1/2} = \sqrt{(1-\rho_1^2)}F^{n-1} + \rho_1\delta^{n+1/2}$$

      with $\delta^{n+1/2} \sim \mathcal{N}(0, C_1)$

      Sample $r \sim U(0,1)$

      Set $\theta^{n+1/2} := (F^{n+1/2}, D^{n-1}, Ds^{n-1})$

      If $r < \alpha(\theta^{n+1/2}, \theta^n)$

      $$F^n = F^{n+1/2}$$

Else

$$F^n = F^{n-1}$$

Update $\theta^n := (F^n, D^{n-1}, Ds^{n-1})$

b) Sample the proposal of $D$

$$D^{n+1/2} = \sqrt{(1+\rho_2^2)}D^{n-1} + \rho_2\delta^{n+1/2}$$

with $\delta^{n+1/2} \sim \mathcal{N}(0, \mathcal{C}_2)$

Sample $r \sim U(0,1)$

Set $\theta^{n+1/2} := \left(F^n, D^{n+1/2}, Ds^{n-1}\right)$.

If $r < \alpha\left(\theta^{n+1/2}, \theta^n\right)$

$$D^n = D^{n+1/2}$$

Else

$$D^n = D^{n-1}$$

a) Sample the proposal of $Ds$

$$Ds^{n+1/2} = \sqrt{1-\rho_3^2}Ds^{n-1} + \rho_3\delta^{n+1/2}$$

with $\delta^{n+1/2} \sim \mathcal{N}(0, \mathcal{C}_3)$

Sample $r \sim U(0,1)$

Set $\theta^{n+1/2} := \left(F^n, D^n, Ds^{n+1/2}\right)$.

If $r < \alpha\left(\theta^{n+1/2}, \theta^n\right)$

$$Ds^n = Ds^{n+1/2}$$

Else

$$Ds^n = Ds^{n-1}$$

Update $\theta^n := (F^n, D^n, Ds^n)$

End

## III. DATA ACQUISITION, NUMERICAL IMPLEMENTATION AND EVALUATION

### A. DATA ACQUISITION

For this study, we used data from 21 subjects (35+/- 6.4 years old) recruited between October 2020 and March 2023 under an ethically approved protocol named DIANE (Dépistage par Irm des Anomalies d'adhésioN placentairE, NCT04328532). The average gestational age was 33 weeks + 2 days. MR imaging data were obtained using a 3T clinical scanner (MAGNETOM Prisma, Siemens Healthcare, Erlangen, Germany). 18 channel body coils with built-in spinal coil elements were used. The whole acquisition was performed in a free-breathing manner, in either the lateral or supine position depending on the patients' comfort. The total acquisition time was within 30 minutes, including both the morphological assessment and functional assessment.

Anatomic $T_2$-weighted (Half-Fourier Acquisition Single-shot Turbo spin Echo sequence) images were obtained in the axial, placental coronal and placental sagittal orientation to cover the whole placenta. The in-plane resolution was 0.89 x 0.89 mm$^2$ and the slice thickness was 5 mm.

Six-minute of IVIM data acquisition was performed in the placenta sagittal plane. Eleven b-values, therefore $N_b = 11$ and b-value = 0, 15, 45, 80, 115, 205, 245.345, 470, 700, 1000 s/$mm^2$ were used. The echo time (TE) and the repetition time (TR) were 60ms and 6000ms, respectively. The in-plane resolution was 1.5625 x 1.5625 mm$^2$ with a 5 mm slice thickness and a1 mm gap between slices.

### B. NUMERICAL IMPLEMENTATION

First, we estimated the $T_2$-weighted anatomical super-resolved reference. We used the implementation proposed in [28] (step 1 in Figure 1).

In the second step, we numerically performed the inter-b-value motion correction (Step 2 in Figure 1). We numerically solved optimization problem (3) using the regularization weight $\lambda_2$= 2.5 and 100 iterations. For each b-value $b_i$, for $i \neq 0$, the $z^{th}$ slice $I_z^{bi}$ within the three-dimensional image obtained is registered to the corresponding $z^{th}$ slice $I_z^{b0}$ within the three-dimensional reference image obtained using the first b-value $b_0$. A 60×60 deformation grid was set on $I_z^{bi}$ for deformable registration.

In the third step, we performed image co-registration (Step 3 in Figure 1). We interpolated the anatomical super-resolved MR image estimated in Step 1 to a grid of the same size as the IVIM image. The latter was used as the reference, and inter-b-value motion-corrected IVIM image was used as the floating image. We solved optimization problem (4) with 50 iterations. The $z^{th}$ slice $I_z^{b0}$ within the three-dimensional inter-b-value motion-corrected IVIM data corresponding to b-value $b_0$ was registered to the corresponding $z^{th}$ slice $I_z^{Anatomic}$ within the three-dimensional anatomic reference image (interpolated image). A 60×60 deformation grid was set on $I_z^{b0}$.

In the fourth step, we implemented Algorithm 1. using maxIter = 5000 (Step 4 in Figure 1), with the first 2000 iterations for the burn-in step of the MCMC method. The step-size parameters $\rho_1$, $\rho_2$ and $\rho_3$ were optimized empirically using a grid search approach. we have identified the following values: $\rho_1$= $\rho_2$ = $\rho_3$ = 0.002 as optimal choice for this application.

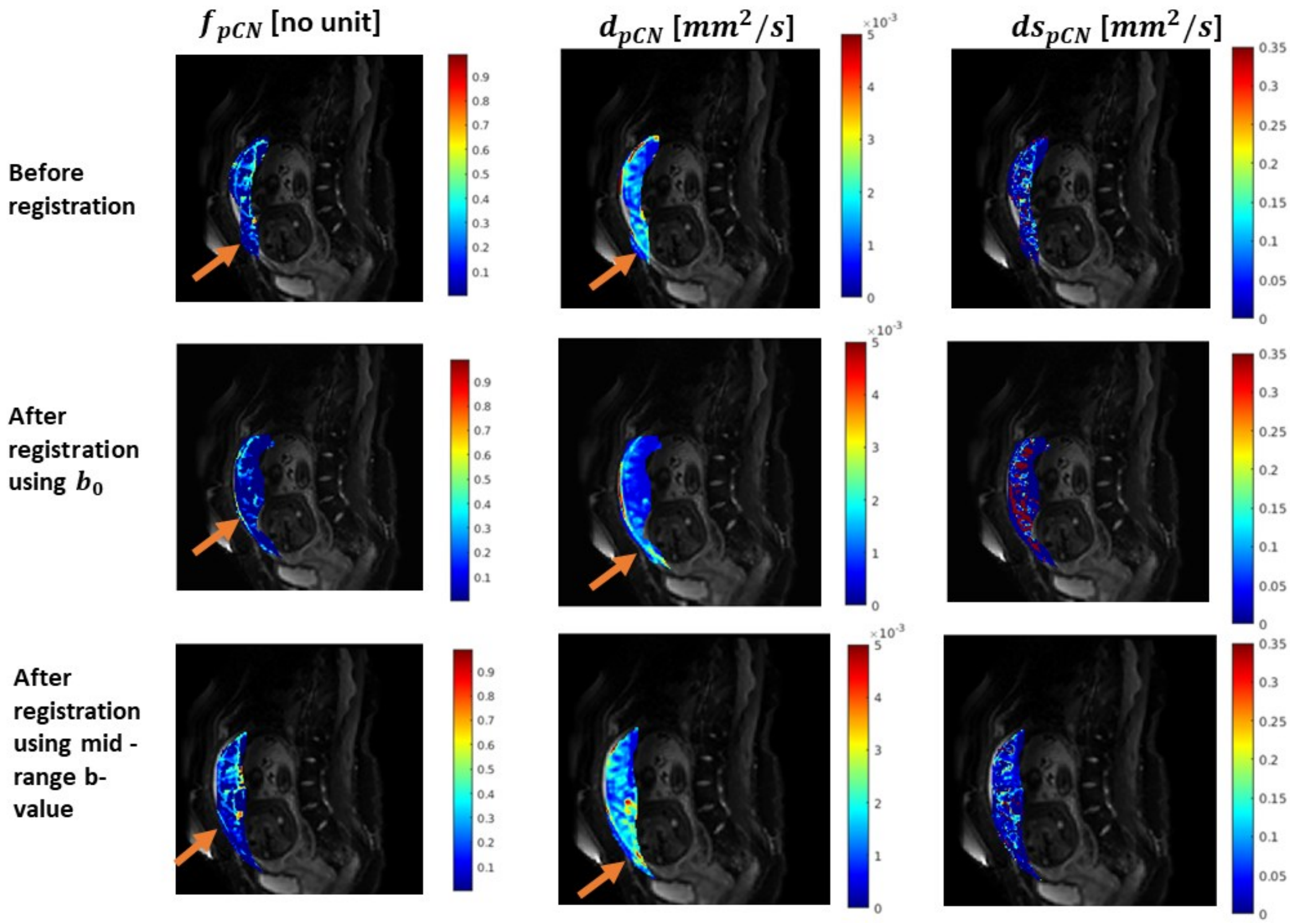

FIGURE 3. IVIM parameters estimates obtained with pCN sampling algorithm before and after motion correction.

We fixed $C_1$ = 0.01, $C_2$ = 0.01 and $C_3$= 0.05. MATLAB 2023a was used for the numerical implementation.
The estimation of the anatomic reference (step 1) has been performed using a computationally efficient algorithm. For Step 2 et 3 concerning inter-b-value motion correction and image co-registration, the computation cost depends on the control point grid used. Finer grid resolution implies higher computational complexity [23]. Concerning IVIM parametric map estimation using the proposed Algorithm 1, at each iteration, the computation of the acceptance rate α costs O(n), where n is the data size, and only the IVIM parameters are sampled (see equation (14)) compared to the conventional Metropolis-Hasting random walk in which additional variables (i.e. mean and covariance matrix of the IVIM parameters, see equation (9)) are also needed to be sampled which add extra computation.

### C. EVALUATION

Two assessments were performed. First, we quantitatively and qualitatively evaluated the impact of the proposed motion correction method on the IVIM parametric maps obtained with the proposed IVIM-pCN estimation method using the first b-value $b_0 := 0\, mm^2/s$ and the mid-range b-value $b_i := 205\, mm^2/s$ as the reference for inter-b-value motion correction.
For quantitative comparison, we used the Mean Absolute Error (MAE) between the observed and fitted data. The middle slice going through placenta with three slices before and after it were selected for an ROI-based evaluation. Averaged based value was obtained in the placenta region of the selected slices for each patient.
In the second part, after motion correction, we compared the proposed IVIM parameter estimation method to the classic Markov chain Monte-Carlo (MCMC) random walk (rw), the segmented Bayesian approach (SegBSP) introduced in [20] and least square Levenberg-Marquadt (LSQ) methods. We calculated the mean value and the standard deviation (std) of placental IVIM parameters obtained with these algorithms over the 21 subjects. Furthermore, we computed the placental Mean Absolute Error (MAE) over the patients.
Markov Chain Monte-Carlo (MCMC) random walk (rw) method was considered as the reference estimation method using the implementation from Orton et al. [11], with 5000 iterations those 2500 iterations for the burn-in step. For the SegBSP method, we used the implementation from [20].

## IV. RESULTS

### A. IMPACT OF MOTION CORRECTION ON IVIM DATA

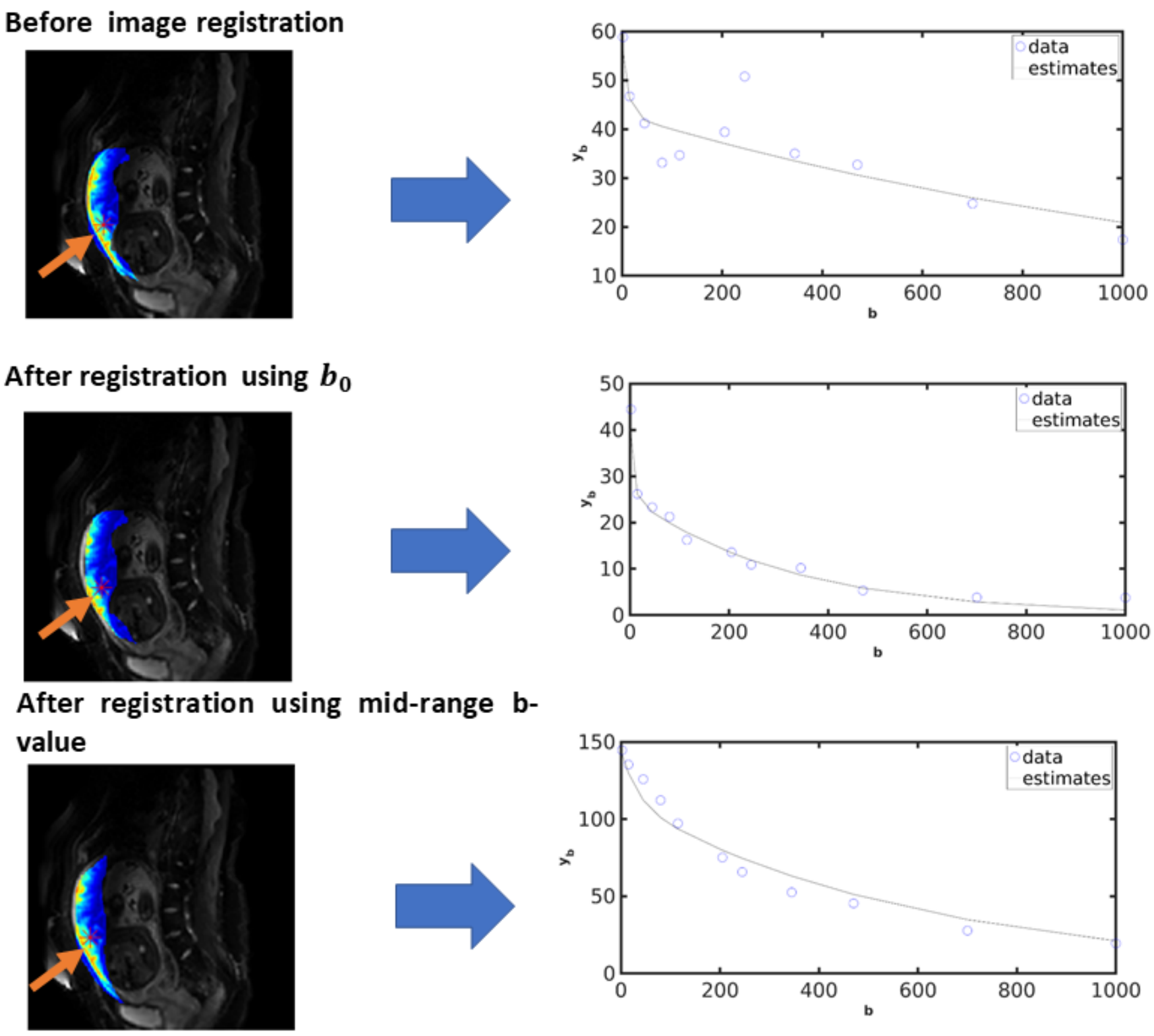

**FIGURE 4.** IVIM fitting using estimates given by pCN sampling algorithm before and after motion correction.

Parametric maps estimated using the proposed IVIM-pCN algorithm before and after using our motion correction method, where the first b-value $b_0$ and the mid-range b-value images are employed as the reference, are shown in Figure 3. Qualitatively, we observed better delineation of perfusion and diffusion boundaries after motion correction (see the orange arrow).

Figure 4 shows an example of IVIM curve fitting at one point located within the placental region before and after motion correction, when the first b-value $b_0$ and the mid-range b-value are set as the reference image for inter-b-value motion correction. The fitting was greatly improved after motion correction for both reference images choices in the inter-b-value motion correction steps. We remark visually that inter-b-value motion correction using $b_0 := 0\, mm^2/s$ provides better fitting.

The median value of the placental perfusion fraction parameter f, when we use the first b-value images $b_0$ as the reference for the inter b-value motion correction, is inferior to that obtained without image registration. The Mean Absolute Error (MAE) range value of the fitting error was reduced after motion correction for both references for the inter-b-value motion correction using the first b-value $b_0$ images and the mid-range b-value images, with a slightly better MAE using the former as the reference. Boxplots of IVIM parameters within the placental region of interest (ROI) (median over slices) for the 21 subjects and the median value of the Mean Absolute Error (MAE) within the placental ROI for the same cohort of 21 subjects are shown in Figure 5. The median value of the placental perfusion, when we use the first b-value image $b_0$ as the reference for the inter b-value motion correction, is inferior to that obtained without image registration. The Mean Absolute Error (MAE) range value of the fitting error was reduction reduced after motion correction for both references for the inter-b-value motion correction using the first b-value $b_0$ image and the mid-range b-value image. These findings are consistent with those presented in [19], with a similar motion correction.

### B. COMPARISON OF IVIM PARAMETERS ESTIMATES AFTER MOTION CORRECTION

In Figure 6, we show one example of IVIM parametric maps using the proposed IVIM estimation method (IVIM-pCN), MCMC random walk (rw), the segmented Bayesian method (SegBSP) and the full non-linear least square (LSQ) method using Levenberg-Marquadt algorithm for one patient, after motion correction where the first b-value $b_0$ is used for the inter-b-value motion correction. patient, after motion correction where the first b-value $b_0$ is used for the inter-b-value motion correction. Whereas the least squares method (LSQ) using Levenberg-Marquadt algorithm provides noisy IVIM parametric estimates, the MCMC random walk, the segmented Bayesian approach (SegBSP) and the proposed IVIM-pCN methods provide similar smoother estimates.

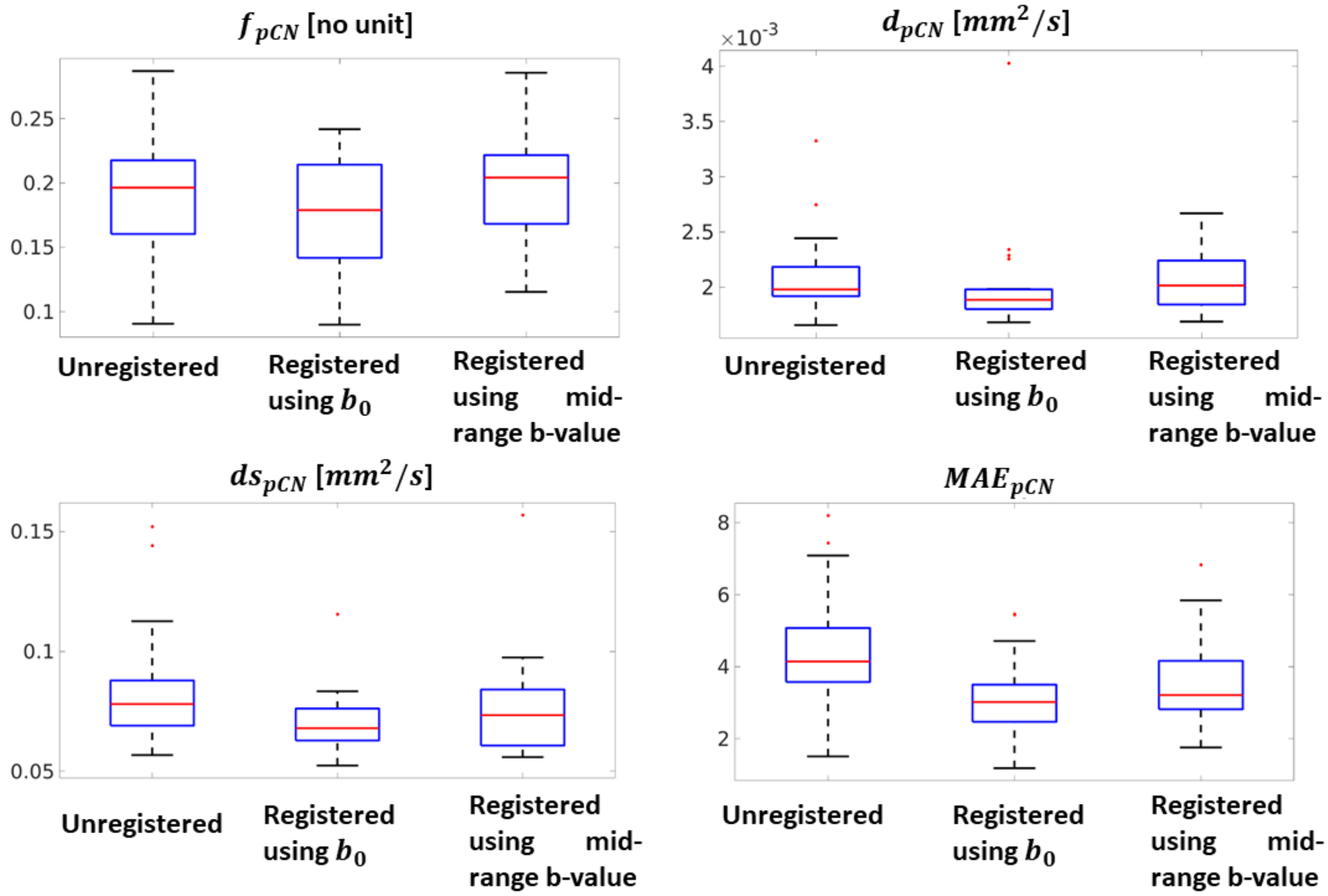

**FIGURE 5.** Boxplot of placental ROI median value of IVIM parameters and Mean Absolute Error (MAE) for all patients.

We also provided example in supplementary material (Figure S1) showing IVIM parameters obtained with the proposed IVIM-pCN estimation method, after applying our motion correction strategy with the mid-range b-value images as the reference for the inter-b-value motion correction.

The standard deviations within the placental region of interest provided by both the IVIM-pCN and MCMC random walk was close, and smaller than those of LSQ. The mean absolute error (MAE) calculated using the pCN and rw methods were also similar. A more complete description of the results can be found in Table 1, with the mean value, standard deviation and MAE calculated in the placenta ROI, with median averaging over slices, and averaging over all 21 subjects, before and after motion correction. One advantage of preconditioned Crank-Nicholson (pCN) sampling method over MCMC random walk is the computation time. Indeed, using the pCN method, we do not estimate the hyperparameters μ and Σ; this reduces the computation time by approximatively 30% to 40% per patient, depending on the data volume size (39% in average).

## V. DISCUSSIONS

In this work, we proposed to optimize placental IVIM parameter estimation by aligning all IVIM data to a super-resolved anatomical volume originally designed for morphological assessment in the placenta exam. We also proposed to accelerate the widely used random walk MCMC method for IVIM parameter estimation using a new sampling strategy called preconditioned Crank-Nicholson sampling.

The proposed framework is assessed using in vivo data. Qualitatively, the results showed that within the framework using the first b-value $b_0$ IVIM images as the reference for the inter-b-value motion correction, the parametric maps show sharper placental boundary than those obtained without motion correction. When using the mid-range b-value images as reference, the extracted parametric maps are also improved compared to those obtained without registration, however not as good as those ones using $b_0$. Quantitatively, we evaluated residual errors before and after image registration and showed a decrease in the MAE metric for the two different references used for the inter-b-value motion correction.

Similar results were obtained using a model-based registration tool for placental MR imaging [19].

TABLE I
RESULTS WITHOUT THE PROPOSED MOTION CORRECTION STRATEGY AND WITH MOTION CORRECTION

| | | **Mean** | | | **Standard deviation** | | | **MAE** |
|---|---|---|---|---|---|---|---|---|
| | | f (no unit) | d ($mm^2$/s) | ds ($mm^2$/s) | f (no unit) | d ($mm^2$/s) | ds ($mm^2$/s) | |
| **pCN** | **Without registration** | 0.1964 | 0.0020 | 0.0780 | 0.1964 | 0.0013 | 0.1797 | 4.1415 |
| | **With Registration using $b_0$** | 0.1789 | 0.0019 | 0.0680 | 0.1934 | 0.0011 | 0.1679 | **3.0169** |
| | **With Registration using the mid-range b-value** | 0.2000 | 0.0020 | 0.0734 | 0.1886 | 0.0012 | 0.1664 | 3.1349 |
| **rw** | **Without registration** | 0.1983 | 0.0019 | 0.0741 | 0.1974 | 0.0013 | 0.1636 | 4.3856 |
| | **With Registration using $b_0$** | 0.1787 | 0.0018 | 0.0672 | 0.1933 | 0.0011 | 0.1679 | **2.9901** |
| | **With Registration using the mid-range b-value** | 0.2038 | 0.0020 | 0.0728 | 0.1887 | 0.0012 | 0.1663 | 3.2107 |
| **LSQ** | **Without registration** | 0.3534 | 0.0018 | 0.2221 | 0.2257 | 0.0015 | 0.3736 | 3.2290 |
| | **With registration using $b_0$** | 0.3387 | 0.0017 | 0.1829 | 0.2232 | 0.0014 | 0.3346 | **2.2822** |
| | **With Registration using the mid-range b-value** | 0.3433 | 0.0019 | 0.1717 | 0.2134 | 0.0014 | 0.3349 | 2.6180 |
| **SegBSP** | **Without registration** | 0.2877 | 0.0015 | 0.0448 | 0.1838 | 0.0006 | 0.0640 | 3.4245 |
| | **With registration using $b_0$** | 0.2772 | 0.0015 | 0.0430 | 0.1887 | 0.0007 | 0.0645 | **2.5399** |
| | **With Registration using the mid-range b-value** | 0.2875 | 0.0016 | 0.0437 | 0.1724 | 0.0006 | 0.0619 | 2.7919 |

Model-based methods were exploited here rather than learning-based methods owing to the small dataset and its robustness in terms of performance.

MR examination of pregnant women is a challenging task, not only because of its multi-motion interruption sources, but also because of the safety and patient comfort constraints leading to other factors, such as limited acquisition time. Therefore, the motion correction strategy of such examinations cannot be adapted to existing methods. Both the MR exam and algorithms need to be designed. Motion can be handled in a prospective manner for diffusion-weighted acquisition as proposed in [18] using navigators to update the slice location. However, such methods lead to longer acquisition times and, therefore, reduce the time required for other examinations. For retrospective motion correction, most published studies have focused on fetal brain data, in which algorithms such as slice-to-volume registration (SVR) and super-resolution reconstruction are widely used to obtain a super-resolved volume. Normally such algorithms need to create a bounding box to restrict the region as an input. Either the fetal brain or the body gains computational advantages as these parts are compact thus bounding boxes can easily contain all the necessary voxels that need to be computed. However, the placenta is a smashed plate that extends over the uterus, especially during the 3rd trimester. Therefore, an intersection-based SVR method was used in this study rather than the commonly used bounding box.

Joint motion correction and super-resolution methods have been proposed to improve quantitative MRI [17]. In the brain, MR structural information and diffusion tensor imaging have also been combined to improve brain connectome modeling and mapping to a standard brain atlas. All of them used only the diffusion data without incorporating high resolution anatomical data [2]. However, these studies were for different purpose and brain fiber tracks were extracted in the end, where potential registration to align to standard atlas may be more important.

The preconditioned Crank-Nicholson (pCN) sampling method was introduced to accelerate the estimation of IVIM placental parameters using a Gaussian prior and Bayesian framework. The results are visually similar to those provided by the MCMC random walk used in previous studies for IVIM parameters estimation [11], but less noisy than to that provided by full non-linear least square (LSQ) Levenberg-Marquadt algorithm. This is consistent with findings of Orton et al. [11]. In contrast, pCN sampling technique is less expensive than the random walk sampling. Its advantage also lies in the fact that the ROI-based prior distribution step is no longer necessary, therefore the estimation pipeline is more automatic. One limitation of (pCN) sampling estimation method is that it considers a normal prior distribution 14], blocking other prior information to be used as in [20]. All the hyperparameters used herein, ρ1, ρ2, ρ3, C1, C2, C3, are arbitrary fixed, which may affect the final results.

Similar motion correction framework has been previously studied. Uss et al. [41] introduced a motion correction framework for three-dimensional $T_2^*$ relaxometry using $T_2$-weighted anatomic MR images. Landmark-based registration was used to reorient $T_2^*$ to the anatomic space. However, the T2* volume used, which is from a fast sequence and intra-measurement motion, can be considered minimal and did not affect the fitting. Therefore, in this work, the registration was between the anatomic volume and reconstructed T2* maps. The motion problem for the IVIM is more complicated.

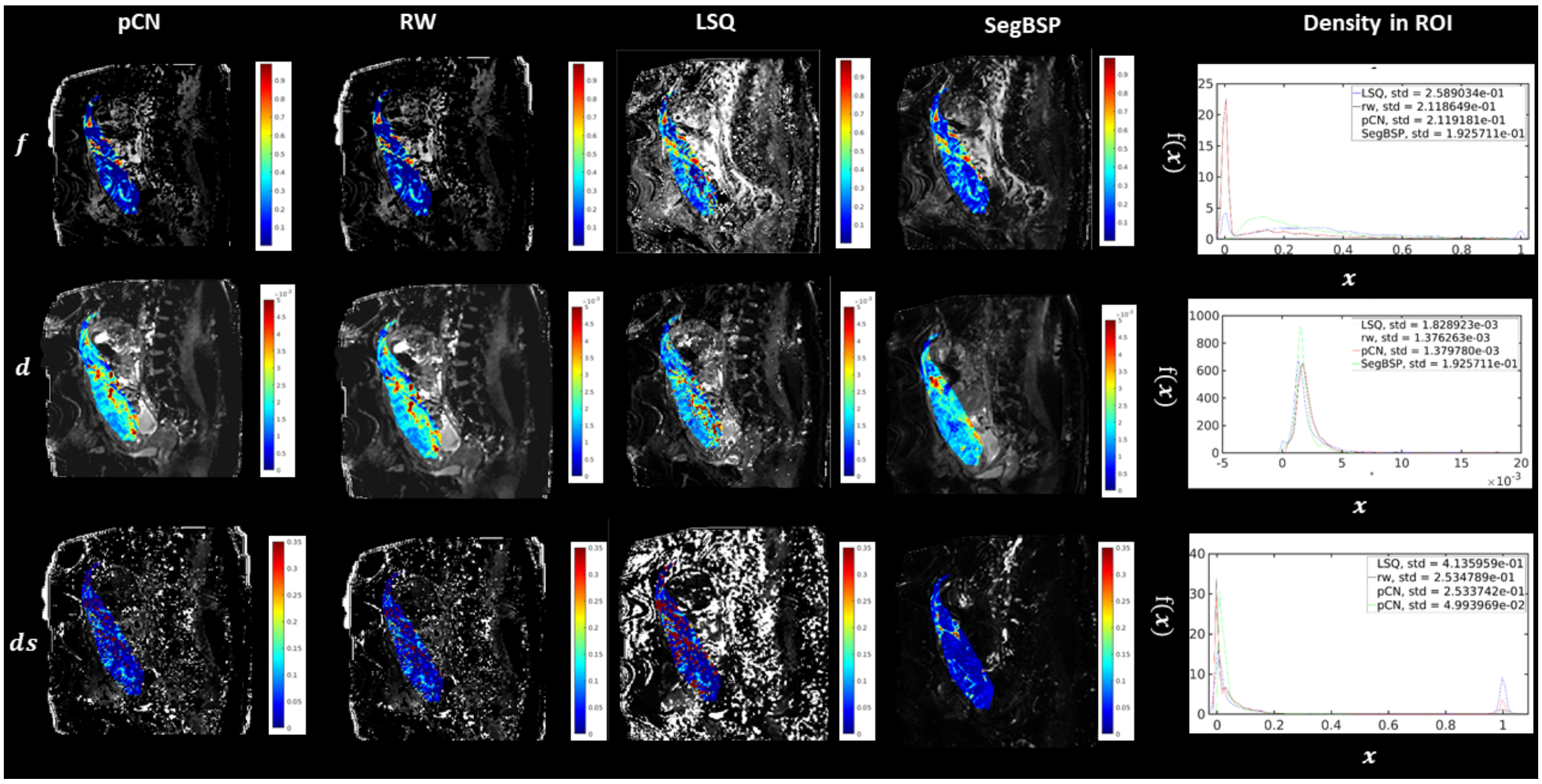

**FIGURE 6.** IVIM parametric maps obtained with IVIM-pCN, rw, SegBSP and LSQ after motion correction using the first b-value as the reference for inter-b-value motion correction, and the probability density for the same data. Red curve is for pCN, black for rw, blue for LSQ and green for SegBSP.

In the literature, most IVIM parameter estimation using Bayesian approach used Metropolis-Hasting random walk method. Spinner et al. [20] introduced a new Bayesian method using the segmented likelihood (SegBSP) and showing improvement of IVIM parameter estimation in cancer over Metropolis-Hasting random walk. Results given by this method show that the later gives the small mean absolute error (MAE) compared to the Metropolis-Hasting sampling (rw) and the proposed method (pCN). However, another metric might give another outcome. In addition, SegBSP is computationally expensive. For instance, the estimation of IVIM parametric maps for one slice last around 795.6 s for SegBSP whereas it takes less than 500s for RW and pCN.

Data employed for this work comes from a clinical protocol with a specific target population. When acquired at higher spatial resolution the computation complexity of IVIM parameters estimation would correspondingly augment however the IVIM resolution of our data is almost at the current technical limitation that one can achieve in clinical protocol for the population of pregnant women with diagnostic quality. Therefore, the proposed algorithm should remain stable for large scale real-world system in terms of scalability.

Data used are from the ongoing clinical protocol DIANE protocol targeting PAS patients, by the time of the work submitted, we only have access to 21 included subjects. We are aware that the small sample size in our study is a limitation. However, this is because PAS is a rare pregnancy condition, and thus our database remains precious, and worth of studying carefully. To include more patients in our study will be our next step in the near future.

The limitations of the proposed framework and estimation methods include the debatable choice of the IVIM image acquired with the first b-value $b_0$ as reference for inter-b-value image registration. The mean b-value image can be adopted to consider the influence of both low and high b-value images [34]. Concerning the co-registration of inter-b-value motion-corrected IVIM data, the reference anatomic MR images used were interpolated using cubic interpolation before registration because it extends the influence of more points [38], but other interpolation kernels can also be used.

## VI. CONCLUSION

In this work, we proposed an innovative framework for analyzing placental IVIM data. The framework used a 3D super-resolved, patient-specific images as an anatomical reference. It included inter- and intra-b-value motion correction, and an accelerated Bayesian estimation of IVIM parametric maps, based on the preconditioned Crank-Nicholson (pCN) sampling method. In a small cohort of 21 patients targeting placenta accrete spectrum, the framework was shown to improve the mean absolute fitting error, and the pCN method was able to speed up parameter estimation compared to other state-of-the-art Bayesian methods. We also acknowledge that one limitation is that this speed-up was at the cost of introducing extra hyperparameters, which may be challenging to optimize. The framework may also be of interest for the study of other moving organs and the estimation of other MR parametric maps such as the relaxation time.

TABLE II
LIST OF SYMBOLS USED

| Symbol | Explanation | Symbol | Explanation |
|---|---|---|---|
| $y_i$, and $\breve{y}_i$ for $i = 1,2,3$ | Multi-slice anatomic $T_2$-weighted images and the motion-corrected in the axial, coronal and sagittal planes, respectively | $D_i$, $B_i$ and $T_i$ | Downsampling, blurring and warping operator, respectively |
| $x$ and $\hat{x}$ | Super-resolved anatomic image and its estimate, respectively | $R_1$ and $\lambda_1$ | Beltrami prior and its weight |
| $\Omega \subset R^2$ | Image domain of anatomic T2-weighted image slice | $b_i$, for $i = 0,2,3,\ldots,10$ | B-value used |
| $N_b$ | The number of b-value used | $I_m^i$ and $I_f^0$ | IVIM slice obtained with b-value $b_i$ for $i \neq 0$ and $b_0$, respectively |
| $t_i$ and $t_i^*$ | Warping affecting the IVIM slice $I_m^i$ and its estimate, respectively | $R_2$ and $\lambda_2$ | Isotropic total variation prior and its weight, respectively |
| $\mathcal{F}$ | Image dissimilarity (local cross-correlation) | $k_i$ and $k_i^*$ | Control points grid and its estimate, respectively |
| $V^0$ | Three-dimensional IVIM data corresponding to $b_0 = 0\ s/mm2$ | $\hat{x}_{interp}^i$ and $I_{interbv}^{i,0}$ | slice of the interpolated anatomic and the corresponding slice in the inter-b-value motion-corrected image associated to the first b-value $b_0$ |
| $y_i^{IVIM}$ and $\varepsilon_i$ | IVIM data and associated noise, respectively | $f, d, ds$ (and F, D, Ds) | IVIM parameters (and their transformed) |
| $\mu$ and $\Sigma$ | the mean of the IVIM parameter and its covariance matrix, respectively | $\pi$ | Sampler function |
| $\theta^n$ and $\theta^{n+1/2}$ | Current state of the chain and the proposal point, respectively | $\rho, \rho_1, \rho_2, \rho_3$ | Step-size of pCN sampling method |
| $\mathcal{F}$ and $\mathcal{E}$ | Starting and end set of $\pi$ | $\alpha$ | Acceptance probability |
| $\delta^{n+1/2}$ | Normally distributed variable linked to sampling | $C, C_1, C_2, C_3$ | Covariance matrix for the sampling |
| $U$ and $\mathcal{N}$ | Uniform and Normal distribution, respectively | $I_z^{bi}$, $I_z^{b0}$ and $I_z^{Anatomic}$ | z-th slice image corresponding to the b-value $b_i, b_0$ and the anatomic image |

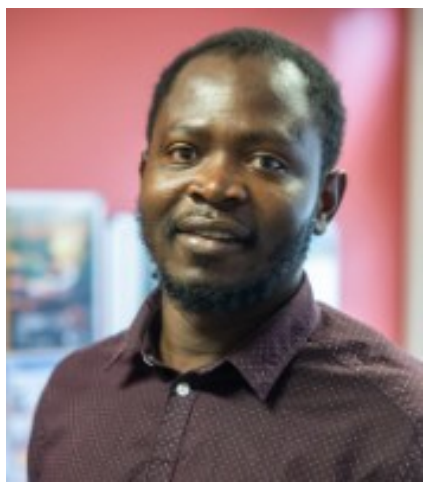

**Mbaimou Auxence NGREMMADJI** received the Bachelor in Mathematics from the Université Cheikh Anta Diop, Senegal, in 2015, and Master in Mathematical modeling from the Université Pierre et Marie Curie, France, in 2019. He is currently a postdoctoral scholar with the IADI laboratory in Nancy.

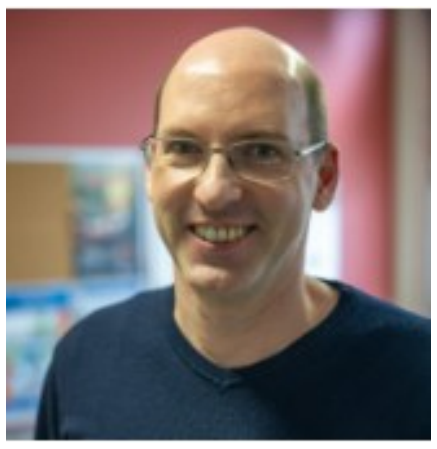

**Freddy ODILLE** received the Ph.D. degree from Nancy University, Nancy, France, in 2007. In 2011, he was appointed as a Research Fellow of Inserm (French Institute of Healthcare Research), IADI Team, Nancy. Since 2015, he has also been with Nancy University Hospital, Nancy. In 2023, he was promoted to research director of Inserm. He is currently the Deputy Head of IADI. His research interests include magnetic resonance imaging, image reconstruction, biomedical signal and image processing, and cardiac electrophysiology modeling and intervention. Since 2022, he has been an elected member of Inserm's scientific panel on healthcare technology.

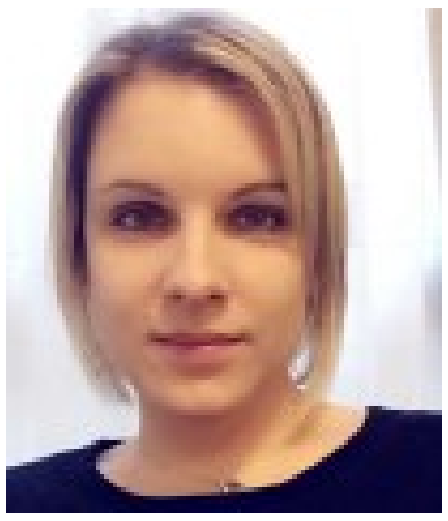

**Charline BERTHODLT** received the M.D. degree in gynecology and obstetrics in 2017, and the Ph.D. degree in 2023, from Université de Lorraine, France. In 2024, she was appointed as a professor at Université de Lorraine and Centre Régional Hospitalier Universitaire (CHRU) of Nancy. She is currently with the maternity of CHRU Nancy. She was principal investigator for numerous clinical studies.

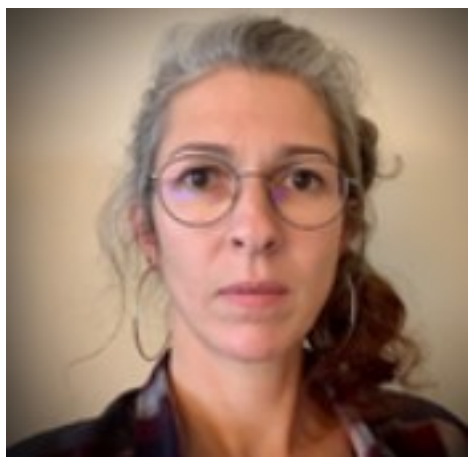

**Marine BEAUMONT** received the Ph.D. degree from the Université Joseph Fourier, Grenoble, France, in 2007. She was postdoctoral researcher at the Sickkids Hospital, in Toronto, Canada. In 2009, she integrated the Centre Hospitalier Regional Universitaire (CHRU) and participated to methodological development of MR imaging for clinical use. She is currently with the Centre d'Investigation Clinique (CIC-IT) in Nancy.

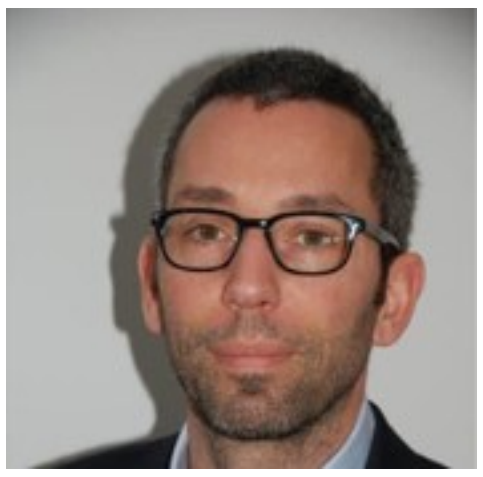

**Olivier MOREL** received M.D. in gynecology and obstetrics from the Université de Lorraine. He was appointed as a professor at Université de Lorraine and Centre Régional Hospitalier Universitaire (CHRU) of Nancy. He is currently the head of the maternity of CHRU Nancy and the general secretary of the Collège National des Gynécologues et Obstétricien Français (CNGOF). He has been a principal investigator of numerous clinical studies.

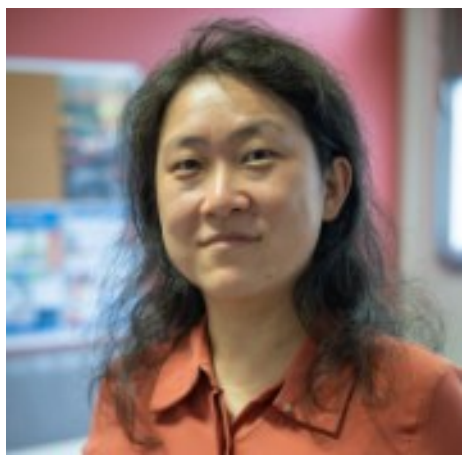

**Bailiang CHEN** received her Ph.D. degree from University College London, UK, in 2013 with an EU Marie Curie funding. After finishing, she started her research career in France by joining the IADI lab and the Centre d'Investigation Clinique (CIC-IT) in Nancy. Her research interests include obstetric magnetic resonance imaging, image reconstruction, biomedical signal and image processing, and multiscale vasculature analysis. She is currently a research engineer. Since 2024, she has been the scientific lead of women imaging research of the IADI team.